\documentclass[a4paper]{llncs}
\usepackage{latexsym,amssymb,upgreek}
\usepackage{xspace}
\usepackage{enumerate}
\usepackage{mathrsfs}
\usepackage{graphics,graphicx}
\usepackage{pstricks,pst-node,pst-tree}
\usepackage{setspace}

\long\def\COMMENT#1{}

\newcommand{\syntree}[1]{\mathit{S}_{#1}}

\newcommand{\Equiv}{\mbox{\bf =}}
\newcommand{\bfeq}{\Equiv}

\newcommand{\myCompa}{\ensuremath{( \bcomp)_a\;}}
\newcommand{\myCompb}{\ensuremath{( \bcomp)_b\;}}
\newcommand{\myNCompa}{\ensuremath{( \uneg\bcomp)_a\;}}
\newcommand{\myNCompb}{\ensuremath{( \uneg\bcomp)_b\;}}

\newcommand{\weight}{{\mathit{weight}}}

\newcommand{\At}{{\mathit{Lit}_{\theta}}}
\newcommand{\Atc}{{\mathit{Lit}_{(\uneg\bcomp)}}}

\newcommand{\NF}{\ensuremath{\; {\mathsf{nf}}}}
\newcommand{\NINT}{\ensuremath{\; {\mathsf{nf}_{\uneg}}}}
\newcommand{\cp}{{\mathsf{cp}}}
\newcommand{\dec}{{\mathsf{dec}}}

\newcommand{\conv}{\mathop{\smallsmile}}

\newcommand{\secondone}{\ensuremath{(\{\one,\cup,\cap\}\bcomp \_)}}

\newcommand{\NBool}[1]{\ensuremath{\mathsf{Bool}_{#1}}\xspace}

\newcommand{\uneg}{\mathop{\textrm{--}}}
\newcommand{\bcomp}{\mathbin{\mbox{\boldmath $;$}}}
\newcommand{\bop}{\mathbin{\sharp}}

\newcommand{\deduction}{deduction }

\newcommand{\one}{{\mathbf{1}}}

\title{A Dual Tableau-based Decision Procedure for a Relational Logic with the Universal Relation (Extended Version)}

\author{Domenico Cantone\inst{1} \and Marianna Nicolosi Asmundo\inst{1} \and \\ Ewa Or{\l}owska\inst{2}}
\institute{ %
{Universit\`a di Catania, Dipartimento di Matematica e Informatica\\
~email:~\texttt{cantone@dmi.unict.it, nicolosi@dmi.unict.it}} \and %
{National Institute of Telecommunications, Warsaw, Poland\\
~email:~\texttt{orlowska@itl.waw.pl}}
}

\begin{document}


\maketitle

\begin{abstract}
We present a first result towards the use of entailment inside relational dual tableau-based decision procedures. To this end, we introduce a fragment of $\mathsf{RL}(\one)$, called $\secondone$, which admits a restricted form of composition, $(R\bcomp S)$ or $(R \bcomp \one)$, where the left subterm $R$ of $(R\bcomp S)$ is only allowed to be either the constant $\one$, or a Boolean term neither containing  the complement operator nor the constant $\one$, while in the case of $(R \bcomp \one)$, $R$ can only be a Boolean term involving relational variables and the operators of intersection and of union.
We prove the decidability of the $\secondone$-fragment by defining a dual tableau-based decision procedure with a suitable blocking mechanism and where the rules to decompose compositional formulae are modified so to deal with the constant $\one$ while preserving termination.

The $\secondone$-fragment properly includes the logics presented in previous work and, therefore, it allows one to express, among others, the multi-modal logic \textsf{K} with union and intersection of accessibility relations, and the description logic $\mathcal{ALC}$ with union and intersection of roles.
\end{abstract}
\section{Introduction}
The relational representation of various non-classical propositional logics has been systematically analyzed in the last decades \cite{Orl}. A uniform relational framework based on the logic of binary relations $\mathsf{RL}(\one)$, presented in \cite{Orl88} and called \emph{relational dual tableau}, showed to be an effective logical means to represent in a modular way three fundamental components of a formal system: its syntax, semantics, and deduction system. Relational systems have been defined for modal and intuitionistic logics, for relevant and many-valued logics, for reasoning in logics of information and data analysis, for reasoning about time and space, etc.

The formalization of non-classical logics in $\mathsf{RL}(\one)$ is based on the fact that once the Kripke-style semantics of the considered logic is known, formulae can be treated as relations. In particular, since in Kripke-style semantics formulae are interpreted as collections of objects, in their relational representation they are seen as right ideal relations. In the
case of binary relations this means that $(R \bcomp \one) = R$ is satisfied,
where `$\bcomp$'  is the composition operation on binary relations and `$\one$' is the universal relation.

One of the most useful features of the relational methodology is that, given a logic with a relational formalization, we can construct its relational dual tableau in a systematic and modular way.

Though the relational logic $\mathsf{RL}(\one)$ is undecidable, it contains
several decidable fragments.  In many cases, however, dual tableau
proof systems are not decision procedures for decidable fragments of
$\mathsf{RL}(\one)$.  This is mainly due to the way decomposition and
specific rules are defined and to the strategy of proof construction.

Over the years, great efforts have been spent to construct dual tableau
proof systems for various logics known to be decidable; little
care has been taken, however, to design dual tableau-based decision
procedures for them.  On the other hand, it is well known that when a proof
system is designed and implemented, it is important to have decision
procedures for decidable logics.  In \cite{FoNi06}, for
example, an optimized relational dual tableau for $\mathsf{RL}(\one)$,
based on Binary Decision Graphs, has been implemented.  However, such
an implementation turns out not to be effective for
decidable fragments.

As far as we know, relational dual tableau-based decision procedures can be found in \cite{Orl} for fragments of $\mathsf{RL}(\one)$ corresponding to the class of first-order formulae in prenex normal form with universal quantifiers only, in \cite{GoMu11a,GoMu11b} for the relational logic corresponding to the modal logic $\mathsf{K}$, in \cite{CanNicOrl10,CanNicOrl11} for fragments of $\mathsf{RL}$ characterized by some restrictions in terms of type $(R \bcomp S)$, in \cite{GHM13} for a class of relational logics admitting a single relational constant with the properties of reflexivity, transitivity, and heredity, and in \cite{CanGolNic14} for a class of relational fragments extending the ones introduced in \cite{GHM13} by allowing a countable infinity of relational constants with the properties of reflexivity, transitivity, and heredity.

Throughout the paper the terms of type $(R \bcomp S)$ will be referred to as compositional terms. Similarly, the formulae built with compositional terms will be referred to as compositional formulae.

In some cases, like in \cite{GHM13} and in \cite{CanGolNic14}, fragments with relational constants satisfying some fixed properties are considered. Therefore, dual tableau-based decision procedures are endowed with specific rules to treat relational constants and their properties. The design of specific rules  often needs much care because termination of the proof procedure must be guaranteed. This task is delicate especially when the proof system provides several specific rules for different relational constants, and when the relational constants are related to each other.

An alternative way to treat properties of relational constants and of relational variables, and of relations between them, is to use \emph{relational entailment}. Relational entailment can be formalized in the logic $\mathsf{RL}(\one)$ as follows. Given relations $R,R_1,\ldots,R_n$, with $n \geq 1$, one has that $R_1 \bfeq \one,\ldots,R_n\bfeq\one$ imply $R \bfeq \one$ in a model if and only if $(\one \bcomp (\uneg(R_1\cap\ldots\cap R_n))\bcomp \one) \cup R \bfeq \one$ holds. As a consequence of that, any validity checker for $\mathsf{RL}(\one)$ can also be applied to verification of entailment.

Introduction of entailment inside relational proof systems permits to eliminate specific rules and, consequently, to keep the set of decomposition rules small.
In its relational formalization, however, entailment involves the universal constant $\one$ on the left hand side and on the right hand side of compositional terms. Thus, the design of a relational dual tableau-based decision procedure where entailment is admitted is a challenging task that requires special care.

In this paper we present a first result towards the use of entailment inside relational dual tableau-based decision procedures. We introduce a fragment of $\mathsf{RL}(\one)$, called $\secondone$, admitting a restricted form of composition where the left subterm, $R$, of any term of type $(R\bcomp S)$ is allowed to be either the constant $\one$ or any term constructed from the relational variables by applying only the operators of relational intersection and union. Similarly, terms of type $(R \bcomp \one)$ are admitted only if $R$ is a Boolean term neither containing  the complement operator nor the constant $\one$.

We prove that the $\secondone$-fragment is decidable by defining a dual tableau-based decision procedure where a suitable blocking mechanism has been introduced and rules for compositional and complemented compositional formulae have been appropriately modified to deal with the constant $\one$ while preserving termination.

This fragment properly includes the logics presented in \cite{CanNicOrl10} and, therefore, it can express the multi-modal logic \textsf{K} with union and intersection of accessibility relations, and the description logic $\mathcal{ALC}$ with union and intersection of roles. Furthemore it can express via entailment
properties of the form `$ \mathsf{r} \subseteq \uneg (\mathsf{s_1} \cup \mathsf{s_2})$' and `$(\mathsf{s_1} \cup \mathsf{s_2})\subseteq \uneg  \mathsf{r}$', where $\mathsf{r}, \mathsf{s_1}$, and $\mathsf{s_2}$ are relational variables.

The paper is organized as follows. In Sect.~\ref{sec:logic} we briefly review the syntax and semantics of the relational logic $\mathsf{RL}(\one)$ together with its dual tableau. In Sect.~\ref{sec:tools} we introduce some useful notions which will be used in the rest of the paper. In Sect.~\ref{sec:secondone} we present the $\secondone$-fragment, its dual tableau-based decision procedure, and prove termination and correctness of the latter. Finally, in Sect.~\ref{sec:conclusions}, we draw our conclusions and give some hints for future work.

\section{The Relational Logic $\mathsf{RL}(\one)$ and its Dual Tableau}\label{sec:logic}

In this section we review the logic
$\mathsf{RL}(\one)$ and its dual tableau in full extent (see also
\cite{CanNicOrl11} and \cite{Orl}).

Let $\mathbb{RV}$ be a countably infinite set of \emph{relational
variables} $\mathsf{p},\mathsf{q},\mathsf{r},\mathsf{s},\dots$ and
let $\one$ be a \emph{relational constant}.  Then, the set
$\mathbb{RT}$ of \emph{relational terms} of $\mathsf{RL}(\one)$ is the
smallest set of terms (with respect to inclusion) containing all
relational variables and the relational constant $\one$, and which is
closed with respect to the \emph{relational operators}  `$\cap$',
`$\cup$', `$\bcomp $' (binary) and `$\uneg$', `$\phantom{\cdot}^{\conv}$' (unary).

Let $\mathbb{OV}$ be a countably infinite set of \emph{object
(individual) variables} $x,y,z,w,\ldots$.  Then,
$\mathsf{RL}(\one)$-formulae have the form $xRy$, where $x,y \in
\mathbb{OV}$ and $R \in \mathbb{RT}$.  $\mathsf{RL}(\one)$-formulae of
type $x\one y$ and $x\mathsf{r}y$, with $\mathsf{r} \in \mathbb{RV}$,
are called \emph{atomic} $\mathsf{RL}(\one)$-formulae.  A
\emph{literal} is either an atomic formula or its complementation
(namely a formula of type $x(\uneg\one) y$ or $x (\uneg\mathsf{r})
y$).  For a relational operator `$\sharp$' other than `$\uneg$', by a
($\sharp$)-\emph{term} we mean a relational term whose main operator is
`$\sharp$', and by a $(\uneg\sharp)$-\emph{term} we indicate a relational term
having `$\uneg$' followed by `$\sharp$' as its main operator. A ($\sharp$)-\emph{formula}
(resp., $(\uneg\sharp)$-\emph{formula}) is a formula whose relational term is a ($\sharp$)-\emph{term}
(resp., $(\uneg\sharp)$-\emph{term}).  A
\emph{Boolean term} is a relational term involving only the Boolean
operators `$\uneg$', `$\cup$', and `$\cap$'.

$\mathsf{RL}(\one)$-formulae are interpreted in $\mathsf{RL}(\one)$-\emph{models}. An $\mathsf{RL}(\one)$-model is a structure $\mathcal{M} = (U,m)$, where $U$ is a nonempty universe and $m : \mathbb{RV} \rightarrow \wp(U \times U)$ is a given map which is homomorphically extended to the whole collection $\mathbb{RT}$ of relational terms as follows:
\begin{itemize}
\item $m(\one) = U \times U$;
\item $m(\uneg R) = (U \times U) \setminus m(R)$;
\item $m(R \cup S) = m(R) \cup m(S)$;
\item $m(R \cap S) = m(R) \cap m(S)$;
\item $m(R\bcomp S) = m(R)\bcomp m(S)$

      $\phantom{m(R\bcomp S)} = \{(a,b)\hspace*{-0.05cm} \in\hspace*{-0.05cm} U\times U \hspace*{-0.05cm}:\hspace*{-0.05cm}(a,c)\hspace*{-0.05cm}\in\hspace*{-0.05cm} m(R) \mbox{ and }  (c,b)\hspace*{-0.05cm} \in\hspace*{-0.05cm}
m(S), \mbox{ for some $c \in U$}\}$;
\item $m(R^{\conv}) = (m(R))^{\conv} = \{(b,a) \in U\times U : (a,b)\hspace*{-0.05cm}\in \hspace*{-0.05cm}m(R)\}$.
\end{itemize}
Let $\mathcal{M} = (U,m)$ be an $\mathsf{RL}(\one)$-model. A
\emph{valuation} in $\mathcal{M}$ is any function $v :
\mathbb{OV} \rightarrow U$. Given an object variable $z$ in $\mathbb{OV}$,  a valuation $v_1$ is a  $z$-\emph{variant} of another valuation $v$ if $v_1(x) = v(x)$, for every $x \in \mathbb{OV}$ such that $x \neq z$.
\emph{Satisfaction} of an
$\mathsf{RL}(\one)$-formula $xRy$ by an $\mathsf{RL}(\one)$-model $\mathcal{M} = (U,m)$
and by a valuation $v$ in $\mathcal{M}$ is defined by
$$\mathcal{M},v \models xRy ~\mbox{ iff }~ (v(x),v(y)) \in m(R).$$
An $\mathsf{RL}(\one)$-formula $xRy$ is \emph{true} in a model $\mathcal{M} = (U,m)$ if $\mathcal{M},v \models xRy$, for every valuation $v$ in $\mathcal{M}$.
An $\mathsf{RL}(\one)$-formula $xRy$ is said to be \emph{valid} if it is \emph{true} in all
$\mathsf{RL}(\one)$-models.
An $\mathsf{RL}(\one)$-formula $xRy$ is \emph{falsified} by a
model $\mathcal{M} = (U,m)$ and by a valuation $v$ in $\mathcal{M}$ if $\mathcal{M},v \not\models xRy$.
It is \emph{falsifiable} if there are a model
$\mathcal{M}$ and a valuation $v$ in $\mathcal{M}$ such that $\mathcal{M},v \not\models xRy$.  An $\mathsf{RL}(\one)$-set is a finite set of $\mathsf{RL}(\one)$-formulae $\{\varphi_1,\ldots,\varphi_n\}$ such that for every $\mathsf{RL}(\one)$-model $\mathcal{M}$ and for every valuation $v$ in $\mathcal{M}$ there exists an $i \in \{1,\ldots,n\}$ such that $\varphi_i$ is satisfied by $v$ in $\mathcal{M}$. Plainly, the first-order disjunction of the formulae in an $\mathsf{RL}(\one)$-set is valid in first-order logic.

Proof development in dual tableaux proceeds by systematically
decomposing the (disjunction of the) formula(e) to be proved till a
validity condition is detected, expressed in terms of axiomatic sets
(see below).  Such an analytic approach is similar to Beth's tableau
method, with the difference that the two systems work in a dual manner.
Duality between tableaux and dual tableaux has been analyzed in depth in
\cite{GolOrl07}.

$\mathsf{RL}(\one)$-dual tableaux consist of decomposition rules, which
allow one to analyze the structure of the formula to be proved valid,
and of axiomatic sets, which specify the closure conditions.  The
decomposition rules for $\mathsf{RL}(\one)$ are listed in Table
\ref{decomprulesRLF}.  In these rules, `$,$' is interpreted as
disjunction and `$|$' as conjunction. A rule is $\mathsf{RL}(\one)$-correct whenever the premise is an $\mathsf{RL}(\one)$-set if and only if each of its consequents is an $\mathsf{RL}(\one)$-set. The rules presented in Table
\ref{decomprulesRLF} are proved $\mathsf{RL}(\one)$-correct in \cite{Orl}.


\begin{table}[bt]
\caption{\label{decomprulesRLF} $\mathsf{RL}(\one)$ decomposition
rules.}
\centerline{\begin{tabular}{ccccc}
 $(\cup)$ &
$\begin{array}{c}
  x (R \cup S) y
\\\hline
  x R y, x S y
\end{array}$ &~~~&
$(\uneg\cup)$ & $\begin{array}{c}
  x (\uneg(R \cup S)) y
\\\hline
  x (\uneg R) y\; |\; x (\uneg S) y
\end{array}$\\[.4cm]
$(\cap)$ & $\begin{array}{c}
  x (R \cap S) y
\\\hline
  x R y| x S y
\end{array}$ &~~~&
$(\uneg\cap)$ & $\begin{array}{c}
  x (\uneg(R \cap S)) y
\\\hline
  x (\uneg R) y, x (\uneg S) y
\end{array}$\\[.4cm]
$(\uneg\uneg)$ & $\begin{array}{c}
  x (\uneg\uneg R) y
\\\hline
  x R y
\end{array}$ \\[.4cm]
$(^{\conv})$ & $\begin{array}{c}
  x (R^{\conv}) y
\\\hline
  y R x
\end{array}$ &~~~&
$(\uneg\;^{\conv})$ & $\begin{array}{c}
  x (\uneg (R^{\conv})) y
\\\hline
  y (\uneg R) x
\end{array}$\\[.4cm]
$(\bcomp )$ & $\begin{array}{c}
  x (R\bcomp S) y
\\\hline
  x R z, x(R\bcomp S)y \;|\; zSy, x(R\bcomp S)y
\end{array}$
&~~~&
$(\uneg \bcomp )$ & $\begin{array}{c}
  x (\uneg(R\bcomp S)) y
\\\hline
  x (\uneg R) z, z(\uneg S)y
\end{array}$\\
 & \mbox{(\small{$z$ is any object variable})}  & & &
 \mbox{(\small{$z$ is a new object variable})}
\end{tabular}}
\end{table}
An $\mathsf{RL}(\one)$-axiomatic set is any set of $\mathsf{RL}(\one)$-formulae containing
a subset of one of the following forms:
\begin{description}
\item [(Ax 1)]$\{xRy, x(\uneg R)y\}$
\item [(Ax 2)]$\{x\one y\}$.
\end{description}
Clearly, an $\mathsf{RL}(\one)$-axiomatic set is also an $\mathsf{RL}(\one)$-set.

Let $xPy$ be an $\mathsf{RL}(\one)$-formula. An \emph{$\mathsf{RL}(\one)$-proof tree} for $xPy$ is
an ordered tree whose
nodes are labelled by disjunctive sets of formulae such that the following properties are satisfied:
\begin{itemize}
\item  the root is labelled with the formula $xPy$;

\item each node, except the root, is obtained from its predecessor
node by an application of a decomposition rule in Table
\ref{decomprulesRLF} to one of the formulae labelling it;

\item a node does not have successors (i.e., it is a leaf node)
whenever its set of formulae is an axiomatic set or none of the rules
of Table \ref{decomprulesRLF} can be applied to its set of formulae.
\end{itemize}
A \emph{branch} $\theta$ of a proof tree is any of its maximal paths; we denote with $\bigcup \theta$ the set of all the formulae contained in the nodes of $\theta$, and with $W_{\theta}$ the collection of object variables occurring in the formulae contained in the nodes of $\theta$.
A node of an $\mathsf{RL}(\one)$-proof tree is \emph{closed} if its associated set of formulae is an axiomatic set.
A branch is closed if one of its nodes is closed.
A proof tree is closed if all of its branches are closed. An $\mathsf{RL}(\one)$-formula is $\mathsf{RL}(\one)$-provable if there is a closed $\mathsf{RL}(\one)$-proof tree for it, referred to as an $\mathsf{RL}(\one)$-\emph{proof}.

A node of an $\mathsf{RL}(\one)$-proof tree is \emph{falsified} by
a model $\mathcal{M} = (U,m)$ and a valuation $v$ in $\mathcal{M}$ if every
formula $xRy$ in its set of formulae is falsified by $\mathcal{M}$
and $v$. A node is \emph{falsifiable} if there exist a model
$\mathcal{M}$ and a valuation $v$ in $\mathcal{M}$ which falsify it.

Correctness and completeness of the $\mathsf{RL}(\one)$-dual tableau
are proved in \cite{Orl}.  However, the logic $\mathsf{RL}(\one)$ is
undecidable.  This follows from the undecidability of the equational
theory of representable relation algebras discussed in
\cite{TarskiGivant87}.

\section{Useful Notions and Properties}\label{sec:tools}
In this section we introduce constructions and notions needed for the presentation of the results of the paper.

\noindent
Let $P$ be any relational term in $\mathsf{RL}(\one)$. The following identities hold:
$$
\begin{array}{|lll|}
  \hline
  (\one \cup P) \equiv (P \cup \one) \equiv \one & \hspace*{1cm} & ((\uneg\one) \cup P) \equiv (P \cup (\uneg \one)) \equiv P\\
  (\one \cap P) \equiv (P \cap \one) \equiv P &  \hspace*{1cm}& ((\uneg\one) \cap P) \equiv (P \cap (\uneg \one)) \equiv (\uneg \one) \\
  (\uneg (\uneg \one)) \equiv \one & & \\
  \hline
\end{array}
$$
Let $H$ be a relational term in $\mathsf{RL}(\one)$ and let $H'$ be obtained from $H$ by systematically simplifying $H$ by means of the above identities. If the simplification is carried out in an inside-out way, the computational complexity of the transformation of $H$ into $H'$ is linear in the length of $H$. Moreover, the following lemma holds.
\begin{lemma}
Let $H$ be a relational term and let $H'$ be constructed as outlined above. Then every Boolean subterm $P$ of $H'$ either is equal to $\one$, or it is equal to $\uneg\one$, or it does not contain $\one$.
\end{lemma}
\begin{proof}
Let $P$ be a Boolean subterm of $H'$. The proof is by induction over the structure of $P$. We distinguish the following cases:
\begin{itemize}
\item If $P \in (\{\one, \uneg\one\} \cup \mathbb{RV})$, then the thesis is trivially satisfied.
\item If $P = (Q \cup S)$ or $P = (Q \cap S)$, then, by the construction of $H'$, both $Q$ and $S$ must be distinct from $\one$ and $\uneg \one$. Moreover, by the inductive hypothesis, they cannot contain $\one$ and, consequently, $P$ cannot contain $\one$ too.
\item If $P = (\uneg Q)$, where $Q \neq \one$, then $Q \neq \uneg \one$ by the construction of $H'$ and the thesis follows by the inductive hypothesis.\qed
\end{itemize}
\end{proof}
It is easy to check that $m(H) = m(H')$ holds for every $\mathsf{RL}(\one)$-model $\mathcal{M} = (U,m)$ and for every $H \in \mathsf{RL}(\one)$. Therefore we can restrict our interest to relational terms simplified as described above.

\subsubsection{Parsing trees.}
As with formulae of standard first-order logic, it is possible to associate to each relational term $P$ of $\mathsf{RL}(\one)$ a
\emph{parsing tree} $\syntree{P}$, which is an ordered tree constructed in the usual way.
Let $\syntree{P}$ be the parsing tree for $P$, and let $\nu$ be a node of $\syntree{P}$. We say that a relational term $Q$ \emph{occurs} within $P$ at position $\nu$ if the subtree of $\syntree{P}$ rooted at $\nu$ is identical to $\syntree{Q}$.
In this case we refer to $\nu$ as an \emph{occurrence} of $Q$ in $P$ and to the path from the root of $\syntree{P}$ to $\nu$ as its \emph{occurrence path}.

An occurrence of a relational term $Q$ within a relational term $P$ is
\emph{positive} if its occurrence path deprived of its last node
contains an even number of nodes labelled with $\{\uneg\}$.
Otherwise, the occurrence is said to be \emph{negative}.

\subsubsection{Normal forms and term components.}\label{sec:procprelim}
Next we define a complement normal form for Boolean relational terms, the
notions of $\NBool{N}$-formula, of $\NBool{}$-construction from $N$,  where $N$ is a set of formulae, and of set
of components of a relational term.  

To begin with, we define  recursively
the function \NINT\
(\emph{complement normal form}) on the set of Boolean relational terms
as follows:
\begin{itemize}
\item $\NINT(\one)$ = $\one$, and $\NINT(\mathsf{s})$ = $\mathsf{s}$, for any relational variable $\mathsf{s}$;
\item $\NINT(S \mathbin{\sharp} H) = \NINT(S) \mathbin{\sharp} \NINT(H)$, for $\sharp\in \{\cap,\cup\}$;
\item $\NINT(\uneg (S \cap H)) = \NINT(\uneg S) \cup \NINT(\uneg H)$;
\item $\NINT(\uneg (S \cup H)) = \NINT(\uneg S) \cap \NINT(\uneg H)$;
\item $\NINT(\uneg \uneg S) = \NINT(S)$.
\end{itemize}

Observe that the  complement
normal form of a term is obtained by successive applications of the De
Morgan laws and of
the law of double negation.

 A term is in complement normal form
whenever the occurrences of the complement operator in it act
only on relational variables or constants.

Plainly, for every Boolean relational term R, the formulae $xR\,y$ and $x\NINT(R)\;y$ are \emph{logically equivalent}, that
is $\mathcal{M},v \models xRy$ if and only if $\mathcal{M},v \models
x\NINT(R)y$, for every model ${\mathcal M}= (U,m)$ and every valuation
$v$ in ${\mathcal M}$.

Let $N$ be a set of formulae, and let $R$, $S$ be two Boolean relational terms. We define the notion of $\NBool{N}$-formulae as follows:
\begin{itemize}
\item every literal $x R y$ in $N$ is a $\NBool{N}$-formula;

\item every formula of the form $x(R \cap S) y$ is a
$\NBool{N}$-formula, provided that either $xRy$ is a $\NBool{N}$-formula and
$S$ is in the complement normal form or $xSy$ is a $\NBool{N}$-formula and
$R$ is in the complement normal form;

\item every formula of the form $x(R \cup S) y$ is a $\NBool{N}$-formula if both $xRy$ and $xSy$ are $\NBool{N}$-formulae.
\end{itemize}
Clearly, if $xSy$ is a $\NBool{N}$-formula, then $xSy$ is syntactically
equal to $x\NINT(S)\;y$ and we write $xSy = x\NINT(S)\;y$.  We say that a
formula $xRy$ has a $\NBool{}$-construction from $N$ if $x\NINT(R)\;y$ is a
$\NBool{N}$-formula.

For example, given a set of formulae $N = \{x(\uneg \mathsf{r})z, x \mathsf{s} z, x(\uneg \mathsf{p})y, z(\mathsf{p} \cup \mathsf{s})y\}$, we have that
the formula $x(((\uneg \mathsf{r}) \cup \mathsf{s}) \cap \mathsf{q})z$ is a $\NBool{N}$-formula because $x ((\uneg \mathsf{r}) \cup \mathsf{s}) z$ is a
$\NBool{N}$-formula and $x \mathsf{q} z$ is in the complement normal form. On the other hand the formula $x(\mathsf{s} \cap (\uneg (\mathsf{q} \cup \mathsf{p})))z$ is not a $\NBool{N}$-formula because $x (\uneg(\mathsf{q} \cap \mathsf{p}))z$ is not in the complement normal form. Both formulae, however, have a $\NBool{}$-construction from $N$ because $x(((\uneg \mathsf{r}) \cup \mathsf{s}) \cap \mathsf{q})z$ is a $\NBool{N}$-formula and
$x\NINT{(\mathsf{s} \cap (\uneg (\mathsf{q} \cup \mathsf{p})))}z = x (\mathsf{s} \cap ((\uneg \mathsf{q}) \cap (\uneg \mathsf{p}))) z$ is a $\NBool{N}$-formula. In this latter case, specifically, $x \mathsf{s} z$ is a $\NBool{N}$-formula and $x ((\uneg \mathsf{q}) \cap (\uneg \mathsf{p})) z$ is in the complement normal form, although it is not a $\NBool{N}$-formula.

Given a term $R$ in $\mathbb{RT}$, an object variable $x$, and a set
of formulae $N$, we define $V(R,x,N)$ as the set of object variables
$z$ such that $xRz$ has a $\NBool{}$-construction from
$N$.  

Let $P$ be a term in $\mathbb{RT}$. We define recursively the set $\cp(P)$ of \emph{the components of the term} $P$ as follows:
\begin{itemize}
\item if $P$ is the relational constant $\one$, or a relational variable, or their complement, then $\cp(P) =\{P\}$;

\item if $P = \uneg\uneg B$, then $\cp(P) = \{P\} \cup \cp(B)$;

\item if $P = B^{\conv}$, then $\cp(P) = \{P\} \cup \cp(B)$;

\item if $P = B \mathbin{\bop} C$ (resp., $P = \uneg (B \mathbin{\bop} C)$), then $\cp(P)
= \{P\} \cup \cp(B) \cup \cp(C)$ (resp., $\cp(P) = \{P\} \cup
\cp(\uneg B) \cup \cp(\uneg C)$), for every binary relational operator
$\bop$.
\end{itemize}
Clearly $\cp(P)$ is finite, for any relational term $P$.

\section{The Fragment \secondone\ and its Decision Procedure}\label{sec:secondone}

Formulae of the fragment \secondone\ of
$\mathsf{RL}(\one)$ are characterized by the fact
that the left subterm $R$ of any term of type $(R\bcomp S)$ in them is allowed
to be either the constant $\one$ or
a term constructed from the relational
variables of $\mathbb{RV}$ by applying only the `$\cup$' and `$\cap$' operators, whereas  the right subterm $S$ of
$(R\bcomp S)$ can involve all the relational operators of
$\mathsf{RL}(\one)$ but the converse
operator `$\phantom{\cdot}^{\conv}$'.

Formally, the set $\mathbb{RT}_{{\secondone}}$ of the terms allowed in \secondone-formulae is the smallest set of terms
containing the constant $\one$ and the variables in $\mathbb{RV}$, and such that if $P, Q, B, H \in \mathbb{RT}_{{\secondone}}$ and $S \in \{H, \one\}$, with
\begin{itemize}
\item $B$ a Boolean term neither containing  the constant $\one$ nor the complement operator, and
\item $H$ containing the constant $\one$ only inside terms of type $(B \bcomp \one)$,
\end{itemize}
then $(\uneg P), (P \cup Q), (P \cap Q), (B \bcomp  S), (\one \bcomp S)  \in \mathbb{RT}_{{\secondone}}$.

\smallskip

\noindent Examples of formulae of the \secondone-fragment are: $x(\uneg((\mathsf{r}_1 \cup \mathsf{s}) \bcomp (\mathsf{p} \bcomp \one)))y$,
$x (\one \bcomp ((\mathsf{r}_1 \cup \mathsf{s}) \bcomp \uneg (((\mathsf{q} \cup \mathsf{p}) \cap \mathsf{r}_1) \bcomp \one)))y$, and
$x (\one \bcomp (((\mathsf{r}_1 \cup \mathsf{s}) \cap \mathsf{r}_2) \bcomp \one))y$.
The latter formula can be rewritten as $x (\one \bcomp (\uneg (\uneg (\mathsf{r}_1 \cup \mathsf{s}) \cup \uneg \mathsf{r}_2) \bcomp \one))y$, where
$(\uneg (\mathsf{r}_1 \cup \mathsf{s}) \cup \uneg \mathsf{r}_2)$ is a relational term formalizing the property `$(\mathsf{r}_1) \cup \mathsf{s}\subseteq \uneg \mathsf{r}_2$'.

\begin{table}[bt]
\caption{\label{decomprulesSECOND} Decomposition
rules proper of the \secondone-fragment.}
\centerline{\begin{tabular}{ccccc}
 $\myCompa$ &
$\begin{array}{c}
    x (B \bcomp S) y
\\\hline
zSy, x(B\bcomp S)y,
\end{array}$ &~~~&
$\myCompb$ & $\begin{array}{c}
  x (\one \bcomp S) y
\\\hline
zSy, x(\one \bcomp S)y
\end{array}$\\[.4cm]
 & \mbox{{\small ($z$ object variable in $V(\uneg B,x, \bigcup \theta)$)}}  & & &
 \mbox{{\small ($z$ is any object variable)}}\\\\
$\myNCompa$ & $\begin{array}{c}
 x \uneg (B \bcomp \one) y
\\\hline
x(\uneg B) z
\end{array}$ &~~~&
$\myNCompb$ & $\begin{array}{c}
   x \uneg (\one \bcomp S) y
\\\hline
z(\uneg S) y
\end{array}$\\[.4cm]
 & \mbox{{\small ($z$ is a new object variable)}}  & & &
 \mbox{{\small ($z$ is a new object variable)}}
\end{tabular}}
\end{table}

The decomposition rules for Boolean formulae of our dual tableau-based decision procedure are just the ones in Table \ref{decomprulesRLF}. Concerning the rule to decompose $(\bcomp )$-formulae, it is convenient to distinguish
between $(\bcomp )$-formulae of type $x (B \bcomp S) y$ and of type $x(\one \bcomp S) y$.  The rule for $(\bcomp )$-formulae of type $x(B \bcomp S) y$ is the
\myCompa-rule of Table \ref{decomprulesSECOND}. There,
$z$ is an object variable belonging to $V(\uneg B,x, \bigcup \theta)$.
Notice that if $S = \one$, the node resulting from the decomposition step is axiomatic. In
case of $(\bcomp )$-formulae of type $x (\one \bcomp S) y$ we apply the rule \myCompb depicted in Table \ref{decomprulesSECOND}.
The variable $z$ used in rule \myCompb is any variable on the current node,
provided that the current branch does not already
contain the formula $zSy$.  Otherwise, $x(\one \bcomp S) y$ is not decomposed with $z$. If $S = \one$, the consideration made with rule \myCompa for the node resulting from the decomposition step holds here as well.

For what concerns $(\uneg \bcomp)$-formulae, we consider first the case of formulae of type $x \uneg(B \bcomp
S)y$.  If $S \neq \one$, such formulae are decomposed by means of the $(\uneg
\bcomp)$-rule in Table \ref{decomprulesRLF}. Otherwise, when $S = \one$ we use the rule \myNCompa
of Table \ref{decomprulesSECOND}.
%
%
%
In the case of formulae of type $x \uneg
(\one \bcomp S) y$, with $S \neq \one$, we use instead the rule \myNCompb of Table \ref{decomprulesSECOND},
with $z$ an object variable new for the current node. The rule is applied
provided that the current branch does not contain any
formula of the form $z'(\uneg S)y$, for any `new' variable $z'$ (otherwise, the formula $x \uneg (\one \bcomp S)
y$ cannot be decomposed). The formula $x (\uneg (\one \bcomp \one))y$ is not decomposed.

 Some remarks on the rules \myCompb, \myNCompa, and \myNCompb of Table \ref{decomprulesSECOND} are in order. Observe that for every $\mathsf{RL}(\one)$-model $\mathcal{M} = (U,m)$, $m(\one) = U \times U$ and thus, for every valuation $v$ and object variables $x$ and $z$, we have $\mathcal{M}, v\models x \one z$ and $\mathcal{M}, v\not\models x (\uneg \one) z$. Thus, we shall assume without loss of generality that each node of any dual tableau for formulae of the \secondone-fragment contains implicitly all literals of type $x (\uneg \one) z$. This accounts for the fact that the decomposition rules \myNCompa and \myNCompb
do not introduce $z (\uneg \one) y$ and $x' (\uneg \one) z$, respectively, on the new node, and rule \myCompb restricts $z$ to be any variable on the current node, rather than any possible variable. However, we shall prove that such a restriction preserves the completeness of the procedure.

It is convenient to introduce the notion of \emph{\deduction tree} for $\mathsf{RL}(\one)$-formulae to give a step-by-step description
of the proof tree construction process.

 As proof trees, \deduction trees are ordered trees whose nodes are labelled with disjunctive sets. However, deduction trees may have some leaf nodes that do not contain any axiomatic set and such that decomposition rules can still be applied to them. As it will be clarified below, \deduction trees can be seen as ``approximations'' of proof trees with the property that they can be completed to proof trees.

\begin{definition}\label{dualtableauM}
Let $x P y$ be a \secondone-formula.  A \emph{\deduction tree} $\mathcal{T}$ for $x P
y$ is recursively defined as follows:
\begin{itemize}
\item [(a)] the tree with only one node labelled with $\{x P y\}$ is a
\deduction tree for $x P y$ (initial \deduction tree);

\item [(b)] let $\mathcal{T}$ be a \deduction tree for $x P y$ and let
$\theta$ be a branch of $\mathcal{T}$ whose leaf node $N$ does not
contain an axiomatic set.\footnote{From now on, we identify nodes with
the (disjunctive) sets labelling them.} The tree obtained from
$\mathcal{T}$ by applying to $N$ either one of the decomposition rules
in Table \ref{decomprulesRLF} (for Boolean formulae and for $(\uneg \bcomp)$-formulae of type $x' \uneg(B \bcomp S)y$, with $S \neq \one$), or one of the
decomposition rules in Table \ref{decomprulesSECOND} (for $(\bcomp)$-formulae, and for $(\uneg \bcomp)$-formulae of type $x' \uneg(B \bcomp \one)y$ and of type $x \uneg(\one \bcomp S)y$)
is a \deduction tree for $x P y$. More precisely, rules applications are
described as follows:
\begin{itemize}
\item if a formula $x' Q y$ occurs in $N$ and a rule with
a single conclusion set of formulae $\Gamma$ (resp., a branching rule
with the conclusion sets $\Gamma_1$ and $\Gamma_2$) is applicable to
$x' Q y$, then we append the node $N' = (N \setminus \{x' Q y\})\cup
\Gamma$ as the successor of $N$ in $\theta$ (resp., the node $N_1' =
(N \setminus \{x' Q y\})  \cup \Gamma_1$ as the left successor of $N$
and the node $N_2' = (N \setminus \{x' Q y\})  \cup \Gamma_2$ as the right
successor of $N$ in $\theta$).
\end{itemize}
\end{itemize}
\end{definition}
Given a branch $\theta$ of a deduction tree, each object variable in $W_{\theta} \setminus\{x,y\}$ is generated by an application of a $(\uneg \bcomp)$-decomposition rule.
We say that a variable $w$ is an \emph{ancestor of degree} $n$ of a variable $z \in W_{\theta}\setminus \{x,y\}$ if there is a sequence $z_1,\ldots,z_n$ of variables in $W_{\theta}\setminus \{x,y\}$, with $z_n = z$ and $n \geq 1$, such that $z_1$ is generated by a $(\uneg \bcomp)$-formula $w(\uneg (B_0\bcomp S_0))y$, $z_2$ is generated by a $(\uneg \bcomp)$-formula $z_1(\uneg (B_1\bcomp S_1))y$,..., $z_n$ is generated by a $(\uneg \bcomp)$-formula $z_{n-1}(\uneg (B_{n-1}\bcomp S_{n-1}))y$, where $w(\uneg (B_0\bcomp S_0))y$, $z_1(\uneg (B_1\bcomp S_1))y$,..., $z_{n-1}(\uneg (B_{n-1}\bcomp S_{n-1}))y$ are formulae of $\theta$. In such a case, we say that $z_1$ is a \emph{descendant of degree} $1$ of $w$ and that $z_n = z$ is a descendant of degree $n$ of $w$.

It is useful to introduce a total order among object variables in $W_{\theta}$, denoted $<_{\theta}$, such that:
\begin{itemize}
\item $x <_{\theta} w$, for every $w \in W_{\theta} \setminus\{x\}$,
\item $x_1 <_{\theta} x_2$, for every $x_1, x_2 \in W_{\theta} \setminus\{x,y\}$ such that $x_1$ has been introduced in $\theta$ before $x_2$,
\item $y <_{\theta} z$, for every $z$ descendant of $y$,
\item $w <_{\theta} y$, for every $w$ that is not a descendant of $y$.
\end{itemize}

\begin{remark}
Notice that the relationship ancestor/descendant is based on the literals of type $x'(\uneg \mathsf{r})z$ that are generated by applying either the $(\uneg\bcomp)$-rule of Table \ref{decomprulesRLF} or the \myNCompa-rule of Table \ref{decomprulesSECOND}, and some consequent Boolean decompositions. Any variable $z$ resulting from the decomposition of a $(\uneg \bcomp)$-formula of type $x (\uneg (\one \bcomp S))y$ is not a descendant of $x$. However, according to the definition of the order $<_{\theta}$, $x <_{\theta} z$ holds.
\end{remark}
Let $\theta$ be a branch of a deduction tree, and let $z(\uneg (B\bcomp S))y$ and $z'(\uneg (B\bcomp S))y$ be two $(\uneg\bcomp)$-formulae occurring in $\theta$. We say that $z'(\uneg (B\bcomp S))y$ \emph{blocks} $z(\uneg (B\bcomp S))y$ (and that $z(\uneg (B\bcomp S))y$ \emph{is blocked by} $z'(\uneg (B\bcomp S))y$), if the following conditions are satisfied:
\begin{itemize}
\item $z(\uneg (B\bcomp S))y$ and $z'(\uneg (B\bcomp S))y$ are identical with the exception of the left object variable,
\item $z'(\uneg (B\bcomp S))y$ has been already decomposed in $\theta$ using the object variable $w$,
\item for every $(\bcomp)$-formula $z (B_1 \bcomp Q)y$ occurring in $\theta$ such that $z (\uneg B_1)w$ has a \NBool{}-construction from the set of literals that result from the Boolean decomposition of $z (\uneg B)w$, the $(\bcomp)$-formula $z'(B_1 \bcomp Q)y$ occurs in $\theta$ as well.
\end{itemize}


\subsection{The Decision Procedure}\label{sec:decproc}

Starting with an initial deduction tree $\mathcal{T}_0$ for a given formula $xPy$, the following procedure constructs a proof tree for $xPy$.

\begin{enumerate}
\item For every non-axiomatic branch $\theta$ of the current deduction tree,
\item  while $\theta$ is non-axiomatic and is further expandable, let $z$ be the smallest variable w.r.t. $<_{\theta}$ such that formulae on $\theta$ with left variable $z$ have not been decomposed in $\theta$. Apply to the formulae on $\theta$ having left variable $z$ the decomposition rules in the following order: Boolean rules, $(\uneg \bcomp)$-rules, rule \myCompa, and then  apply rule \myCompb to decompose the $(\bcomp)$-formulae of type $x(\one\bcomp S)y$ in $\theta$ with the variable $z$ till saturation. We require that:
\begin{itemize}
     \item [a.]all the rules can be applied at most once with the same premise;
      \item [b.] every formula of type $(\uneg\bcomp)$, $z(\uneg (B\bcomp S))y$ is not decomposed provided that it is blocked by a $(\uneg\bcomp)$-formula $z'(\uneg (B\bcomp S))y$ occurring in $\theta$.

         If $z'(\uneg (B\bcomp S))y$ was decomposed in $\theta$ with the variable $w$, then for every literal $z' (\uneg \mathsf{r}) w \in \bigcup \theta$ (obtained from the application of the Boolean rules to $z'(\uneg B)w$) we store the literal $z (\uneg \mathsf{r}) w$ in $\Atc$, a set (empty at the beginning of the execution of the procedure) collecting literals not explicitly occurring in $\theta$ that are needed to construct the model $\mathcal{M}_{\theta}$ (see step 4).
\end{itemize}
\item If the branch $\theta$ is axiomatic and all the other branches on the current deduction tree are axiomatic, then the current deduction tree is a proof tree for $xPy$ and we terminate. Otherwise, if the branch $\theta$ is axiomatic and there are still non-axiomatic branches on the current deduction tree, return to step 1.
\item Otherwise, if $\theta$ is non-axiomatic, namely it is a non-axiomatic not further expandable branch,  we construct from $\theta$ the model $\mathcal{M}_{\theta} = (U_{\theta}, m_{\theta})$ defined as follows. We put $U_{\theta} = W_{\theta}$. Next, let $\At$ be the set of all literals occurring in $\theta$, and let $\Atc$ be defined as in step 2.
    We define the interpretation $m_{\theta}$ by putting $(x',y') \notin m_{\theta}(R)$ if and only if $x' R y' \in (\At\cup \Atc)$.
    Let $v_\theta : \mathbb{OV} \rightarrow U_\theta$ be a valuation such that $v_\theta(x) =_{Def} x$, for every $x \in U_{\theta}$.
    We terminate returning $\theta$, $\mathcal{M}_{\theta}$, and $v_{\theta}$.
\end{enumerate}

The next lemma states two useful properties of the formulae occurring on the deduction trees constructed by the proof procedure above.

\begin{lemma}\label{lemma:term1first}
Let $\mathcal{T}$ be a deduction tree for $xPy$ constructed by an execution of the procedure described above. If $x'Rx''$ is a formula of a branch $\theta$ of $\mathcal{T}$, then
\begin{itemize}
\item [(i)] $R \in \cp(P)$,
\item [(ii)] if $R$ contains the composition operator, then $x'' = y$.
\end{itemize}
\end{lemma}
\begin{proof}
Let $\mathcal{T}_0, \ldots, \mathcal{T}_n$ be a sequence of deduction trees constructed by an execution of the proof procedure illustrated above, where $\mathcal{T}_0 = \{xPy\}$ and each $\mathcal{T}_{i+1}$ is obtained from $\mathcal{T}_i$, for $i = 0,1,\ldots, n-1$ by the application of a decomposition rule and $\mathcal{T}_n = \mathcal{T}$.

We prove by induction on $i$ that, for every $x'Rx''$ occurring on $\mathcal{T}_i$, (i) and (ii) hold. If $i = 0$, $\mathcal{T}_0 = \{xPy\}$ and therefore (i) and (ii) are trivially verified.
 By the inductive step, we assume that (i) and (ii) are true for every formula of $\mathcal{T}_i$. Since $\mathcal{T}_{i+1}$ is obtained from $\mathcal{T}_i$ by the application of a decomposition rule to the leaf node of one of its branches $\bar{\theta}$, all the formulae on $\mathcal{T}_{i+1}$, with the exception of the new formulae originated by the decomposition step, occur also in $\mathcal{T}_i$ and thus satisfy (i) and (ii). Then we have to prove that (i) and (ii)
 hold also for the formulae originating from the decomposition step. %
%
%
%

Let us consider $\bar{\theta}$, its leaf node $\bar{N}$, and the formula $x'Rx''$ chosen to be decomposed. We analyze the possible cases. If $x'Rx''\equiv x'(Q_1 \cap Q_2)x''$, then we append $\bar{N}' = (\bar{N}\setminus \{x'Rx''\}) \cup \{x'Q_1 x''\}$ as the left successor and $\bar{N}'' = (\bar{N}\setminus \{x'Rx''\}) \cup \{x'Q_2 x''\}$  as the right successor of $\bar{N}$. Since $Q_1 ,Q_2 \in \cp(R)$ and $R \in \cp(P)$, we have that $Q_1 ,Q_2 \in \cp(P)$ and thus (i) is satisfied by $x'Q_1 x''$ and $x'Q_2 x''$. If $R$ does not contain the composition operator, (ii) is trivially verified. Otherwise, $x'' = y$, and thus (ii) holds for $x'Q_1 x''$ and for $x'Q_2 x''$ too. The remaining Boolean cases can be proved in an analogous way.

If $x' R x'' = x'(B\bcomp S)x''$ and $z \in V(\uneg B, x', \bigcup\theta)$, we append $\bar{N}' = \bar{N} \cup \{zSx''\}$ as the successor of $\bar{N}$. Since $S \in \cp(R)$ and $R \in \cp(P)$, it follows that $S \in \cp(P)$ and thus (i) holds also for $zSx''$.  Since $x'' = y$, (ii) holds for $zSx''$ as well. The case in which $x' R x'' = x'(\one\bcomp S)x''$ can be treated in an analogous way.

If $x' R x'' = x'(\uneg(B\bcomp S))x''$, with $S \neq \one$, we append $\bar{N}' = (\bar{N} \setminus \{x'(\uneg(B\bcomp S))x''\})\cup \{x'(\uneg B)z, z(\uneg S)x''\}$ as the successor of $\bar{N}$, where $z$ is new for $\bar{N}$. Since $\uneg B, \uneg S \in \cp(R)$ and $R \in \cp(P)$,  $\uneg B, \uneg S \in \cp(P)$ and thus (i) holds for both $x'(\uneg B)z$ and $z(\uneg S)x''$. Since $x'' = y$ and $(\uneg B)$ does not contain the complement operator, (ii) is trivially satisfied for both $x'(\uneg B)z$ and $z(\uneg S)x''$. The cases $x' (\uneg (\one \bcomp S))x''$ and $S = \one$ can be handled in a similar way. \qed
\end{proof}

\subsection{Termination of the procedure}
Let $\mathcal{T}$ be a deduction tree for a formula $xPy$ of the \secondone-fragment constructed according to the procedure described in Sect. \ref{sec:decproc}. To prove that the procedure always terminates it is useful  to give the following characterization of a non-axiomatic not further expandable branch $\theta$ of $\mathcal{T}$:

\begin{itemize}
\item $x'\one y' \notin N$, for every node $N \in \theta$;

\item if $x' R y'$ (resp., $x' (\uneg R) y'$) occurs in a node $N \in
\theta$, then $x' (\uneg R) y'$ (resp., $x' R y'$) does not occur
in any node $N' \in \theta$;

\item if $x' (\uneg \uneg Q) y'$ occurs in a node $N \in \theta$, then there is a node $N' \in
\theta$, successor of $N$, such that $x' Q y' \in N'$;

\item if $x' (Q_1 \cap Q_2) y'$ occurs in a node $N \in \theta$, then there is a node $N' \in
\theta$, successor of $N$, such that either $x' Q_1 y' \in N'$ or $x' Q_2 y' \in N'$;
\item if $x' (Q_1 \cup Q_2) y'$ occurs in a node $N \in \theta$, then there is a node $N' \in
\theta$, successor of $N$, such that $x' Q_1 y' \in N'$ and $x' Q_1 y' \in N'$;
\item if $x' (\uneg (Q_1 \cap Q_2)) y'$  occurs in a node $N \in \theta$, then there is a node $N' \in
\theta$, successor of $N$, such that $x' (\uneg Q_1) y, x' (\uneg Q_2) y' \in N'$;
\item if $x' (\uneg (Q_1 \cup Q_2)) y'$  occurs in a node $N \in \theta$, then there is a node $N' \in
\theta$, successor of $N$, such that either $x' (\uneg Q_1) y'$ or $x' (\uneg Q_2) y' \in N'$;

\item if $x' (B \bcomp S) y$, with $S \neq \one$, occurs in a node $N \in
\theta$, then, for every $z \in V(-B, x', \mathcal{S})$, there is a node $N' \in
\theta$, successor of $N$,
such that $z S y \in N'$;

\item if $x(\one \bcomp S)y$, with $S \neq \one$, occurs in a node $N \in
\theta$, then, for every $w \in W_{\theta}$, there is a node $N' \in
\theta$, successor of $N$,
such that $w S y \in N'$;

\item if $x' (\uneg (B \bcomp S))y \in N$ and is not blocked by any $z'(\uneg (B \bcomp S))y  \in N'$,
with $N'$ predecessor of $N$, then there is a node $N''$, successor of $N$, and an object variable $u$ such that $x'(\uneg B)u, u(\uneg S)y \in N''$, if $S \neq \one$, and $x'(\uneg B)u \in N''$ otherwise;
\item if $x (\uneg (\one \bcomp  S)) y$ occurs in a node $N \in \theta$ and $S \neq \one$, then there is a node $N'\in \theta$ such that $z S y \in
N'$, for some object variable $z$.
\end{itemize}

Next we state and prove some preliminary lemmas and make some remarks.

\begin{lemma}\label{lemma:term2first}
Let $\theta$ be a non-axiomatic not further expandable branch of a deduction tree $\mathcal{T}$ for a formula $xPy$. Then, for every $x' \in W_{\theta}$, $\bigcup \theta$ contains a finite number of formulae of type $x'Rx''$.
\end{lemma}
\begin{proof}
The lemma is immediate for formulae $x'Rx''$ in $\theta$, with $x'' = y$, because $x'$ is fixed and $\cp(P)$ is a finite set. On the other hand, if $x'Rx''$ is in $\theta$ and $x'' \neq y$, then, by Lemma \ref{lemma:term1first} (ii), $R$ is a Boolean term neither containing the complement operator nor the constant $\one$, obviously with $R \in \cp(P)$ by Lemma \ref{lemma:term1first} (i). Thus, by systematic Boolean decomposition, $x'Rx''$ generates inside $\theta$, a finite number of formulae with left variable $x'$ and right variable $x''$. Since $x'$ is fixed and $\cp(P)$ is a finite set, in order to prove that, for any variable $x''\in W_{\theta} \setminus \{x,y\}$, the number of formulae of type $x'Rx''$ in $\theta$ is finite, we have to show that the number of such $x''$s is finite. But this follows from the fact that each right variable of a formula of type $x' R x''$, with $x'' \neq y$, is generated by the decomposition of a $(\uneg\bcomp)$-formula $x' (\uneg(B\bcomp S))y$ and, possibly, by a finite number of Boolean decompositions. By the first part of this lemma, the number of $(\uneg\bcomp)$-formulae $x' (\uneg(B\bcomp S))y$ in $\theta$ is finite, moreover by conditions (a) and (b) on rules application stated in step 2 of the procedure of Sect. \ref{sec:decproc}, each of these formulae can be decomposed at most  once. \qed
\end{proof}

\begin{remark}\label{remark:term2}
Variables generated by $(\uneg \bcomp)$-formulae with left variable $y$ are finitely many because $(\uneg \bcomp)$-formulae of type $y(\uneg(B\bcomp S))y$ are finitely many as well. Moreover these variables are distinct from all the variables generated by the other $(\uneg \bcomp)$-formulae because the $(\uneg \bcomp)$-rule always introduces a new variable when it is applied.
\end{remark}
\begin{remark}\label{remark:term3}
If a variable $w$ is generated by a $(\uneg \bcomp)$-formula $x'(\uneg(B\bcomp S))y$ with $x' \neq y$, then no literal of the form $y (\uneg \mathsf{r}) w$ is in $\theta$. In fact, by Lemma \ref{lemma:term2first} we know that literals of type $y (\uneg \mathsf{r}) z$, with $z \neq y$, are introduced in $N$ only after the decomposition of a $(\uneg \bcomp)$-formula with left variable $y$. But then $z$ cannot be the same variable introduced by a $(\uneg \bcomp)$-formula $x'(\uneg(B\bcomp S))y$ with $x' \neq y$.
\end{remark}
\begin{remark}\label{remark:term4}
Every $(\bcomp)$-formula $w(B\bcomp S)y$ is decomposed only with the variables introduced by the decomposition of $(\uneg \bcomp)$-formulae with left variable $w$ and possibly with the variable $y$.
\end{remark}
\begin{lemma}\label{lemma:term3second}
Every formula $w(B\bcomp S)y$ in $\theta$ can be decomposed a finite number of times.
\end{lemma}
\begin{proof}
Every $(\bcomp)$-formula $w(B\bcomp S)y$ in $\theta$ can be decomposed a number of times equal to the cardinality of the set $V(\uneg B, w, \bigcup\theta)$. By Lemmas \ref{lemma:term1first} and \ref{lemma:term2first}, there are only finitely many $(\uneg \bcomp)$-formulae with left variable $w$. Thus, all the literals with left variable $w$ and right variable $z \neq y$ are finitely many. By Lemma \ref{lemma:term2first} it turns out that literals with left variable $w$ and right variable $y$ are finite too, and therefore the set $V(\uneg B, w, \bigcup\theta)$ must be finite.   \qed
\end{proof}

\begin{lemma}\label{lemma:term6second}
$W_{\theta}$ is finite.
\end{lemma}
\begin{proof}
During the construction of the non-axiomatic not further expandable branch $\theta$, $W_{\theta}$ is enlarged by decomposing formulae of type $(\uneg \bcomp)$. By the conditions on the application of the decomposition rules, each $(\uneg\bcomp)$-formula can be decomposed at most once. Therefore, in order for $W_{\theta}$ to be infinite, $\theta$ must contain an infinite number of formulae of type $(\uneg \bcomp)$. Since $\cp(P)$ is finite, the number of terms of type $(\uneg \bcomp)$ in $\theta$ is finite too. Thus, by Lemma \ref{lemma:term2first} (ii), the only possibility for $\theta$ to contain an infinite number of formulae of type $(\uneg\bcomp)$ is that there is at least one $(\uneg \bcomp)$-formula $x' (\uneg (B\bcomp S))y$ that occurs in $\theta$ infinitely many times, each time with a different left variable. Since, by Lemma \ref{lemma:term3second}, every formula $w(B_1\bcomp S_1)y$ can be decomposed a finite number of times in $\theta$, the formula $x' (\uneg (B\bcomp S))y$, that appears in $\theta$ infinitely many times, each time with a different left variable, must originate from the decomposition of a $(\bcomp)$-formula of type $x(\one \bcomp S_2)y$. By condition (b) of step 2 of the decision procedure in Sect. \ref{sec:decproc}, $x' (\uneg (B\bcomp S))y$ is allowed to appear infinitely often with a different left variable only if there are infinite distinct sets of $(\bcomp)$-terms belonging to $\cp(P)$. But this is not possible because $\cp(P)$ is finite. Thus, the conditions of application of decomposition rules introduced in step 2 guarantee that no $(\uneg \bcomp)$-formula $x' (\uneg (B\bcomp S))y$ is allowed to occur in $\theta$ infinitely many times, each time with a different left variable. Therefore it follows that $W_{\theta}$ is finite.\qed
\end{proof}
Next, we define recursively the $\weight$ of a term by putting:

\vspace*{-0.2cm}

\begin{itemize}
\item $\weight(\mathsf{r}) = \weight(\uneg \mathsf{r}) = \weight(\one) = \weight(\uneg \one)
= 0$;

\item $\weight(A \mathbin{\sharp} P)= \weight(A) + \weight(P) +1$, for $\sharp
\in \{\cup,\cap,\bcomp\}$;

\item $\weight(\uneg (A \mathbin{\sharp} P))= \weight(\uneg A) + \weight(\uneg
P) +1$, for $\sharp \in \{\cup,\cap,\bcomp\}$;

\item $\weight(\uneg \uneg P) = \weight(P) +1$.

%
\end{itemize}

Then we define the weight of a formula $xPy$ as the weight of its term $P$ and the weight of a node as the sum of the weights of the formulae in $N$. In particular, the weight of every $(\bcomp)$-formula and the weight of every $(\uneg \bcomp)$-formula that cannot be decomposed in $N$, according to the decomposition rules and in particular, to the conditions on rules application stated in step 2, is set to 0.
It can be checked that the weight of a node $N$ is 0 if and only if it contains only literals and formulae of types $(\bcomp)$ and $(\uneg \bcomp)$ that cannot be further decomposed, according to the definition of the decomposition rules and of the requirements on rules application in step 2 of the procedure of Sect. \ref{sec:decproc}. Thus, a branch with leaf node of weight 0 is not further expandable.
\begin{lemma}\label{lemma:term7second}
After a finite number of decomposition steps, a branch $\theta$ of a deduction tree for $xPy$ is prolonged to a branch which can be either axiomatic or non-axiomatic and with the weight of the leaf node equal to 0.
\end{lemma}
\begin{proof}
Let $\theta = \theta_1, \theta_2, \ldots$ be such that $\theta_{i+1}$ is obtained from $\theta_i$ by an application of a decomposition rule to the leaf node $N_i$ of $\theta_i$, for $i = 1,\ldots$. If $\theta_i$ results to be an axiomatic branch, then the thesis immediately follows. Otherwise, we reason as shown next.
For every $(\bcomp)$-formula $\varphi$ of $N_i$, of both types $x'(B\bcomp S)y$ and $x(\one \bcomp S)y$, let $\dec(\varphi,N_i)$ be the number of times $\varphi$ has been decomposed on the branch to which $N_i$ belongs. If $\varphi = x(\one \bcomp S)y$, then $\dec(\varphi, N_i) \leq |W_{\theta}|$. By Lemma \ref{lemma:term6second}, $|W_{\theta}|$ is a finite number and once $\dec(x(\one \bcomp S)y,N_i)$ reaches it, $\weight(x(\one \bcomp S)y)$ is set to $0$.
If $\varphi = x'(B\bcomp S)y$, then $\dec(\varphi,N_i)$ is bounded as stated in Lemma \ref{lemma:term3second}.
It turns out that, at each decomposition step, we have either
\begin{enumerate}
\item $\weight(N_i) > \weight(N_{i+1})$, or
\item $\Sigma_{\varphi \in N_i} \dec(\varphi,N_i) < \Sigma_{\varphi \in N_{i+1}} \dec(\varphi,N_{i+1})$.
\end{enumerate}
The first condition holds when the decomposition rule applied to obtain $\theta_{i+1}$ from $\theta_{i}$ is different from the $(\bcomp)$-rule, whereas the second condition holds when the $(\bcomp)$-rule is used.

Since each node contains a finite number of formulae, $\weight(N_i)$ is a nonnegative function and $\dec(\varphi,N_i)$ is bounded for every $(\bcomp)$-formula $\varphi$, after a finite number of steps we obtain a branch $\theta_n$ with a leaf node of weight 0. This means that $\theta_n$ is not further expandable. Moreover, if $\theta_n$ is not closed, then it is a non-axiomatic not further expandable branch. In fact, all the Boolean formulae in $\theta_n$ have been decomposed, all the $(\uneg \bcomp)$-formulae, in view of the conditions of step 2, either have been decomposed into formulae of smaller weight or they have been not decomposed and their weight has been set to 0. Finally, all the $(\bcomp)$-formulae in $\theta_n$ have been decomposed, each finitely many times according to condition (a) of step 2.\qed
\end{proof}
Considering that the procedure of Sect. \ref{sec:decproc} constructs any axiomatic branch and any non-axiomatic not further expandable branch of a proof tree for $x P y$ in a finite number of decomposition steps, we can state the following theorem.
\begin{theorem}[Termination]
The dual tableau procedure for the \secondone-fragment described
in Sect. \ref{sec:decproc} always terminates.
\end{theorem}
\subsection{Correctness of the Procedure}
We show next that the procedure is correct in the sense that when the input formula $xPy$ is valid, the procedure yields a closed (axiomatic) dual tableau for $xPy$, whereas if $xPy$ is not valid the procedure yields a non-axiomatic not further expandable branch $\theta$ of a dual tableau for $xPy$ and a model $\mathcal{M}_{\theta}$ that falsifies every formula on $\theta$ and, in particular, $xPy$ itself.

\begin{lemma}\label{lemma:compl}
Let $\mathcal{T}$ be a \deduction tree for a formula $xPy$ of the $\secondone$-fragment of $\mathsf{RL}(\one)$ constructed as described in the procedure of Sect. \ref{sec:decproc}. If the procedure terminates at step 4 returning a non-axiomatic not further expandable branch $\theta$, a model ${\mathcal M_{\theta}} = (U_{\theta},m_{\theta})$, and a valuation $v_{\theta}$, then ${\mathcal M_{\theta}}$ and $v_{\theta}$ falsify $\theta$.

\end{lemma}
\begin{proof}
We have to prove that ${\mathcal M}_{\theta}$ and $v_{\theta}$ falsify all the formulae in the nodes of $\theta$.
For this purpose,
it is convenient to consider the set $\bigcup \theta$ of all the formulae contained in the nodes of $\theta$.

Let $\varphi$ be a formula of $\bigcup \theta$.  We prove by
induction over the structure of $\varphi$ that ${\mathcal M}_{\theta}$ and $v_{\theta}$
falsify $\varphi$.  The base case is handled as in \cite[Lemma
4]{CanNicOrl11}.  Concerning the inductive step, for simplicity, we
report the proof only for $(\bcomp)$-formulae and for $(\uneg\bcomp)$-formulae.
Let $\varphi = x' (B \bcomp
S)y$, and let us consider first the case where $S \neq \one$.  To prove that ${\mathcal M}_{\theta},v_{\theta}\not\models x' (B \bcomp
S)y$, we
have to show that, for every $z\in W_{\theta} = U_{\theta}$,

\vspace*{-0.3cm}

\begin{equation}\label{eq:lemmacompl1}
\mbox{either }\;{\mathcal M}_{\theta},v_{\theta} \models x' (\uneg B) z \;\;\mbox{ or }
\;\;{\mathcal M}_{\theta},v_{\theta} \models z (\uneg S) y\,.
\end{equation}

\vspace*{-0.1cm}

By a repeated application of the $(\bcomp)$-rule, all the formulae $zSy$, with $z \in V(\uneg B,x',\bigcup\theta)$, occur in $\bigcup\theta$. In particular, each of them belongs to a node of the branch and, by inductive hypothesis, $\mathcal{M}$ and $v$ do not satisfy each of them. Thus, (\ref{eq:lemmacompl1}) is satisfied for every $z \in V(\uneg B,x',\bigcup\theta)$. We have to prove that it is satisfied also for every $z \in (W_{\theta}\setminus V(\uneg B,x',\bigcup\theta))$. We distinguish the case where $x' (\uneg B)z$ has no $\NBool{}$-construction from $\Atc$, and the case where $x' (\uneg B)z$ has a $\NBool{}$-construction from $\Atc$.

Assume first that $z \in (W_{\theta}\setminus V(\uneg B,x',\bigcup\theta))$ and that $x' (\uneg B)z$ has no $\NBool{}$-construction from $\Atc$. We prove that $\mathcal{M},v \models x' (\uneg B) z$ by induction over the structure of $x' (\uneg B) z$.

If $x' (\uneg B) z$ is a literal, then $x' (\uneg B) z \notin \bigcup\theta$. Indeed, if $x' (\uneg B) z \in \bigcup\theta$, then $z$ has to be a member of  $V(\uneg B,x',\bigcup\theta)$ contradicting our hypothesis. Moreover $\Atc$ does not contain $x' (\uneg B) z$ and thus
$\mathcal{M},v \models x' (\uneg B) z$.

If $x' \NF(\uneg B) z = x'(Q_1 \cup Q_2)z$, then at least one of $x'Q_1 z$ and $x'Q_2 z$, say $x'Q_1 z$ (without loss of generality), is not a $\NBool{\bigcup\theta}$-formula, because otherwise $z$ would belong to $V(\uneg B,x',\bigcup\theta)$, and it is not a $\NBool{\Atc}$-formula either, because otherwise $x' (\uneg B)z$ would have a $\NBool{}$-construction from $\Atc$.  Since $x'Q_1 z$ is in complement normal form and it is not a  $\NBool{\Atc}$-formula, it does not have a $\NBool{}$-construction from $\Atc$. Moreover, $z \in (W_{\theta}\setminus V(\uneg Q_1,x',\bigcup\theta))$, because $x'Q_1 z$ is in complement normal form and it is not a $\NBool{\bigcup\theta}$-formula. Thus, by inductive hypothesis, $\mathcal{M},v \models x'Q_1 z$. Hence, $\mathcal{M},v \models x'(Q_1 \cup Q_2)z$, so that $\mathcal{M},v \models x' (\uneg B) z$. Finally, if $x' \NF(\uneg B) z = x'(Q_1 \cap Q_2)z$, then none of $x'Q_1 z$ and $x'Q_2 z$ is either a $\NBool{\bigcup\theta}$-formula, because otherwise $z$ would belong to $V(\uneg B,x',\bigcup\theta)$, or a $\NBool{\Atc}$-formula. Both $x'Q_1 z$ and $x'Q_2 z$ do not have a $\NBool{}$-construction from $\bigcup\theta$ and from $\Atc$, and thus $z \in (W_{\theta}\setminus V(\uneg Q_1,x',\bigcup\theta))$ and $z \in (W_{\theta}\setminus V(\uneg Q_2,x',\bigcup\theta))$. Hence, by inductive hypothesis, $\mathcal{M},v \models x'Q_1 z$ and $\mathcal{M},v \models x'Q_2 z$. Thus $\mathcal{M},v \models x'(Q_1 \cap Q_2)z$, so that $\mathcal{M},v \models x' (\uneg B) z$.

%
%
%

Finally, let us consider the case where $z \in (W_{\theta}\setminus V(\uneg B,x',\bigcup\theta))$ and $x' (\uneg B)z$ has a $\NBool{}$-construction from $\Atc$. By construction of $\Atc$, $\bigcup\theta$ must contain two $(\uneg\bcomp)$-formulae: $x'' \uneg(B_1 \bcomp Q)y$, decomposed in $\theta$ with the variable $z$ and $x' \uneg(B_1 \bcomp Q)y$ not decomposed in $\theta$. In addition, $\bigcup\theta$ contains literals of type $x'' (\uneg \mathsf{r}) z$ differing from the literals of type $x' (\uneg \mathsf{r}) z$ in $\Atc$ only because they contain the variable $x''$ in place of the variable $x'$. The formulae $x'' \uneg(B_1 \bcomp Q)y$ and $x' \uneg(B_1 \bcomp Q)y$ satisfy condition (b) of step 2 of the decision procedure of Sect. \ref{sec:decproc} (in the sense that $x' \uneg(B_1 \bcomp Q)y$ is blocked by $x'' \uneg(B_1 \bcomp Q)y$) and, therefore, by the blocking condition, there must be in $\bigcup \theta$ a $(\bcomp)$-formula $x''(B \bcomp S)y$. Clearly, $z \in V(\uneg B,x'',\bigcup\theta)$, the formula $z S y$ is  in $\bigcup\theta$, and, by inductive hypothesis, it holds that ${\mathcal M}_{\theta},v_{\theta} \not\models z S y$. Thus,  ${\mathcal M}_{\theta},v_{\theta} \models z (\uneg S) y$, as we wished to prove.

Let $\varphi = x' (B \bcomp S)y$, with $S = \one$. To prove that ${\mathcal M}_{\theta},v_{\theta}\not\models x' (B \bcomp
\one)y$, we
have to show that, for every $z\in W_{\theta}$,
$\mathcal{M}_{\theta},v_{\theta} \models x' (\uneg B) z$ holds.
Since $\theta$ is a non-axiomatic not further expandable branch of $\mathcal{T}$, $V(\uneg B,x',\bigcup\theta)$ must be empty. Moreover, $x' (\uneg B)z$ cannot have a $\NBool{}$-construction from $\Atc$. Assume by contradiction that $x' (\uneg B)z$ has a $\NBool{}$-construction from $\Atc$. Then, by reasoning as above, there must be a $(\bcomp)$-formula $x''(B \bcomp S)y$ such that $z \in V(\uneg B,x'',\bigcup\theta)$. Then $z \one y \in \bigcup \theta$, which is a contradiction, since $\theta$ is by hypothesis a non-axiomatic branch. Thus, $x' (\uneg B)z$ does not have a $\NBool{}$-construction from $\Atc$. In addition, it can be shown  that $\mathcal{M}_{\theta},v_{\theta} \models x' (\uneg B) z$ holds, for every $z\in W_{\theta}$. This can be done by induction on the structure of $x' (\uneg B) z$ much along the same lines of a previous case of this proof, where condition (\ref{eq:lemmacompl1}) is verified for every $z \in (W_{\theta}\setminus V(\uneg B,x',\bigcup\theta))$ under the hypothesis that $x' (\uneg B)z$ does not have a $\NBool{}$-construction from $\Atc$.


Let $\varphi = x (\one \bcomp S)y$.
To prove that ${\mathcal M}_{\theta},v_{\theta}\not\models x (\one \bcomp
S)y$, we
have to show that, for every $z\in W_{\theta}$,

\begin{equation}\label{eq:lemmacompl2}
\mbox{either }\;{\mathcal M}_{\theta},v_{\theta} \models x (\uneg \one) z \;\;\mbox{ or }
\;\;{\mathcal M}_{\theta},v_{\theta} \models z (\uneg S) y\, .
\end{equation}
Clearly ${\mathcal M}_{\theta},v_{\theta} \models x \one z$, for every $z \in W_{\theta}$. On the other hand, we know that $z S y \in \bigcup\theta$, for every
$z \in W_{\theta}$, and thus, by inductive hypothesis, we have that
${\mathcal M}_{\theta},v_{\theta} \not\models z S y$. Hence, ${\mathcal M}_{\theta},v_{\theta} \models z (\uneg S) y$ holds, as we wished to prove. Plainly $S \neq \one$ because, by hypothesis, the branch is non-axiomatic.

Let $\varphi = x' (\uneg(B \bcomp S))y$, with $S \neq \one$.  To prove that ${\mathcal M}_{\theta},v_{\theta}\not\models x'(\uneg(B \bcomp
S))y$, we
have to show that there exists a $z\in W_{\theta}$ such that

\vspace*{-0.3cm}

\begin{equation}\label{eq:lemmacompl3}
\mbox{}\;{\mathcal M}_{\theta},v_{\theta} \models x' B z \;\;\mbox{ and }
\;\;{\mathcal M}_{\theta},v_{\theta} \models z  S y\,.
\end{equation}

\vspace*{-0.1cm}

If there is no $(\uneg\bcomp)$-formula $x''(\uneg(B \bcomp S))y$ occurring in $\theta$ that blocks $x'(\uneg(B \bcomp
S))y$, then $x'(\uneg(B \bcomp
S))y$ is decomposed with an object variable $z$, and thus
$x' (\uneg B)z$ and $z (\uneg S)y$ are in $\bigcup\theta$. By inductive hypothesis, ${\mathcal M}_{\theta},v_{\theta} \not\models x'(\uneg B)z$ and ${\mathcal M}_{\theta},v_{\theta} \not\models z (\uneg S) y$ hold. Hence, we have ${\mathcal M}_{\theta},v_{\theta} \models x' B z$ and ${\mathcal M}_{\theta},v_{\theta} \models z  S y$, as we wished to prove. Otherwise, let $x''(\uneg(B \bcomp
S))y$ be a $(\uneg\bcomp)$-formula  blocking $x'(\uneg(B \bcomp
S))y$ and let $w$ be the variable used to decompose $x''(\uneg(B \bcomp
S))y$. Then, $x''(\uneg B) w$ and $w (\uneg S)y$ are in $\bigcup\theta$. By construction of $\mathcal{M}_{\theta}$, we have ${\mathcal M}_{\theta},v_{\theta} \models x' B w$ and, since, by inductive hypothesis, ${\mathcal M}_{\theta},v_{\theta} \not\models w(\uneg S)y$, we have ${\mathcal M}_{\theta},v_{\theta} \models w Sy$ and thus the thesis follows.
The case where $S = \one$ can be treated in a very similar way, considering that ${\mathcal M}_{\theta},v_{\theta}\not\models x'(\uneg(B \bcomp
\one))y$ can be proved showing that there is a $z\in W_{\theta}$ such that ${\mathcal M}_{\theta},v_{\theta} \models x' B z$.

Let $\varphi = x (\uneg(\one \bcomp
S))y$. To prove that ${\mathcal M}_{\theta},v_{\theta}\not\models x(\uneg(\one \bcomp
S))y$, we
have to show that there is a $z\in W_{\theta}$ such that

\vspace*{-0.3cm}

\begin{equation}\label{eq:lemmacompl4}
\mbox{}\;{\mathcal M}_{\theta},v_{\theta} \models x \one z \;\;\mbox{ and }
\;\;{\mathcal M}_{\theta},v_{\theta} \models z  S y\,.
\end{equation}

\vspace*{-0.1cm}

By construction, $\bar{z} (\uneg S) y \in \bigcup\theta$, for some $\bar{z}$, and thus, by inductive hypothesis,
${\mathcal M}_{\theta},v_{\theta} \not\models \bar{z} (\uneg S) y$, and ${\mathcal M}_{\theta},v_{\theta} \models \bar{z} S y$ hold. Clearly, we also have  ${\mathcal M}_{\theta},v_{\theta} \models x \one \bar{z}$, as we wished to prove. If $S = \one$, the thesis is immediate. \qed
\end{proof}

\begin{theorem}
If $xPy$ is a valid formula of the $\secondone$-fragment of $\mathsf{RL}(\one)$, then the procedure described in Sect. \ref{sec:decproc} yields a closed proof tree for $xPy$.
\end{theorem}
\begin{proof} Suppose by way of contradiction that the procedure described in Sect. \ref{sec:decproc} does not yield any
closed proof tree for $xPy$. Then, step 4 is executed and the procedure yields a non-axiomatic not further expandable branch $\theta$, a model $\mathcal{M}_{\theta}$, and a valuation $v_{\theta}$.
By Lemma \ref{lemma:compl}, $\mathcal{M}_{\theta}$ and $v_{\theta}$ falsify $\theta$, namely they falsify each formula on the nodes of $\theta$, and thus,
in particular, $\mathcal{M}_{\theta}, v_{\theta}$ falsify $xPy$ , thus contradicting the hypothesis.
\qed
\end{proof}

Next we show that each decomposition step performed by the decision procedure of Sect. \ref{sec:decproc}  preserves
falsifiability.  This result is needed later in the proof of Theorem
\ref{teo:soundness}.

\begin{lemma}\label{soundlemma}
Let $\theta$ be a branch of a deduction tree for a formula $xPy$ of the \secondone-fragment that is being constructed by the procedure of Sect. \ref{sec:decproc} and let $\theta'$ be obtained from $\theta$ by a decomposition step performed by the decision procedure. If $\theta$ is a falsifiable branch, then $\theta'$ is falsifiable  too.
\end{lemma}
\begin{proof} Assume that $\theta$ is falsifiable and let $\mathcal{M} =
(U,m)$ and $v$ be, respectively, a model and a valuation falsifying
each node of $\theta$.

The branch $\theta'$ is obtained from $\theta$ by decomposing a
non-literal formula $\varphi$ occurring on the leaf node $N$ of
$\theta$ as illustrated in the procedure of Sect. \ref{sec:decproc} and the proof can be carried out according to the type of the
formula $\varphi$.  If $\varphi$ is a Boolean formula, the thesis follows
easily. If $\varphi = x' \uneg(B \bcomp S) y$, with $S \neq \one$ and there is no formula of type $x'' \uneg(B \bcomp S) y$ in $\bigcup\theta$ blocking $\varphi$,  we decompose  $x' \uneg(B \bcomp S) y$ using the $(\uneg\bcomp)$-rule illustrated in Table \ref{decomprulesRLF} and reason as in \cite[Lemma 2(1)]{CanNicOrl11}. Otherwise, if there is a formula of type $x'' \uneg(B \bcomp S) y$ in $\bigcup\theta$  blocking $x' \uneg(B \bcomp S) y$, we do not decompose  $x' \uneg(B \bcomp S) y$. Thus, $N' = N$, $\theta' = \theta$ and trivially $\mathcal{M}, v \not\models N'$ and $\mathcal{M}, v \not\models \theta'$. If  $\varphi = x' \uneg(B \bcomp \one) y$  and there is no formula of type $x'' \uneg(B \bcomp \one) y$ in $\bigcup\theta$  blocking $x' \uneg(B \bcomp \one) y$,  we decompose  $x' \uneg(B \bcomp \one) y$ using the rule \myNCompb of Table \ref{decomprulesSECOND}: $\theta' = \theta N'$, where $N' = (N \setminus \{x' \uneg(B \bcomp \one) y\}) \cup \{x' (\uneg B)z\}$, with $z$ a new variable for $\theta$.
Since $\mathcal{M},v \not\models x' \uneg(B \bcomp \one) y$, then $\mathcal{M},v \models x' (B \bcomp \one) y$, i.e., there is
a $u \in U$ such that $(v(x'),u)\in m(B)$ and $(u,v(y)) \in m(\one)$. Let us consider the $z$-variant $v'$  of $v$, such that $v'(z) = u$. we have  $\mathcal{M},v' \models x' B z$ and $\mathcal{M},v' \models z \one y$. Thus  $\mathcal{M},v' \not\models z(\uneg B) y$ and, since $z$ is a new variable,  $\mathcal{M},v' \not\models N'$. Hence $\mathcal{M},v' \not\models \theta'$, as we wished to prove. The case where there is a formula of type $x'' \uneg(B \bcomp \one) y$ in $\bigcup\theta$  already decomposed, that satisfies  together with $x' \uneg(B \bcomp \one) y$ the conditions on the  applicability of the $(\uneg\bcomp)$-decomposition rule stated in step 2 of the procedure of Sect. \ref{sec:decproc}, can be treated as the previous case.

If $\varphi = x \uneg(\one \bcomp S) y$, then
$\theta' = \theta N'$, where $N' = (N \setminus \{x \uneg(\one \bcomp S) y\}) \cup \{z (\uneg S)y\}$, with $z$ a new variable for $\theta$.
Since $\mathcal{M},v \not\models x \uneg(\one \bcomp S) y$, then $\mathcal{M},v \models x (\one \bcomp S) y$, i.e., there is
a $u \in U$ such that $(v(x),u)\in m(\one)$ and $(u,v(y)) \in m(S)$. Let us consider the $z$-variant $v'$ of $v$,  such that $v'(z) = u$. Then  $\mathcal{M},v' \models x\one z$ and $\mathcal{M},v' \models z S y$ hold. Thus,  $\mathcal{M},v' \not\models z(\uneg S) y$ and since $z$ is a new variable,  we have $\mathcal{M},v' \not\models N'$, so that  $\mathcal{M},v' \not\models \theta'$, as we wished to prove.

If $\varphi = x'(B \bcomp S)y$, the proof can be carried out along the same lines of \cite[Lemma 2(2)]{CanNicOrl11}.

If $\varphi = x(\one \bcomp S)y$ and $z$ is any object variable on $N$ such that $z S y$ is already in $\theta$, we put $\theta' = \theta N'$, with $N' = N \cup \{zSy\}$. The proof that $\mathcal{M},v \not\models \theta'$ can be carried out as for the previous case, namely $\varphi = x'(B \bcomp S)y$.\qed
\end{proof}

We close the section with the following result.

\begin{theorem}\label{teo:soundness}
Let $x P y$ be a non valid relational formula of the $\secondone$-fragment. Then the procedure yields a non-axiomatic not further expandable branch $\theta$ of a dual tableau for $xPy$ and a model $\mathcal{M}_{\theta}$ that falsifies every formula on $\theta$ and, therefore, $xPy$ itself.
\end{theorem}
\begin{proof}
Suppose, by way of contradiction, that the procedure yields a closed proof tree $\mathcal{T}_c$ of $xPy$. By definition, all the branches of $\mathcal{T}_c$
must be closed, namely they must be axiomatic. Let us consider the initial deduction tree $\mathcal{T}_0 = \{xPy\}$. By hypothesis, $\mathcal{T}_c$ is falsifiable. Thus, by iteratively applying Lemma \ref{soundlemma}, $\mathcal{T}_c$ must contain a falsifiable and closed branch $\theta$ which is a contradiction. Thus, the procedure constructs a non-axiomatic not further expandable branch $\theta$ (the first falsifiable branch it encounters) and then it constructs a model $\mathcal{M}_{\theta}$ that, by Lemma \ref{lemma:compl}, falsifies each formula on $\theta$.\qed
\end{proof}

\section{Conclusions and Future Work}\label{sec:conclusions}
Relational entailment allows to deal with properties of relational constants and of relational variables in dual tableau proofs without adding
any specific rule to the basic set of decomposition rules. Using entailment in dual tableau-based decision procedures, however, can be tricky
because the constant $\one$  occurs both  on the left hand side and on the right hand side of composition.

We have presented a dual tableau-based decision procedure for a fragment of the logic $\mathsf{RL}(\one)$ allowing to express some simple forms of inclusion between relations. Specifically, we admit inside entailment only positive occurrences of Boolean terms.
This permits to express inclusion properties of type `$(\mathsf{r}_1) \cup \mathsf{s}\subseteq \uneg \mathsf{r}_2$'.

We plan to extend the expressibility of our relational fragment in order to make entailment widely applicable in dual tableau-based decision procedures. As a first step, we intend to include negative occurrences of Boolean terms inside entailment. In this way we will be able to formulate terms of type $\one \bcomp (\uneg (\uneg (\mathsf{r}_1 \cup \mathsf{s}) \cup \mathsf{r}_2) \bcomp \one)$ expressing the (positive) inclusion property `$(\mathsf{r}_1 \cup \mathsf{s})\subseteq \mathsf{r}_2$'.

Our further aim is to add, inside entailment, some restricted forms of composition so to be able to express terms of type $\one \bcomp (\uneg (\uneg (\mathsf{s} \bcomp \mathsf{s}) \cup \mathsf{s}) \bcomp \one)$ and of type $\one \bcomp (\uneg (\uneg (\mathsf{r} \bcomp \mathsf{r}\bcomp\mathsf{r} ) \cup \mathsf{r}) \bcomp \one)$, stating that the relational variables $\mathsf{s}$ and $\mathsf{r}$ are transitive  (i.e., `$(\mathsf{s}\bcomp\mathsf{s})\subseteq \mathsf{s}$') and tree-transitive (i.e., `$(\mathsf{r}\bcomp\mathsf{r} \bcomp \mathsf{r})\subseteq \mathsf{r}$'), respectively. Being able to express these properties is important if we want to use our dual tableau decision procedure for modal logics to reason with incomplete information \cite{DeOrVa99}.

We also aim at introducing the converse relation `$^{\conv}$' and the identity relation `$\one'$' inside entailment for the purpose of dealing with properties
such as symmetry and reflexivity.

\bigskip

\noindent\textbf{Acknowledgments.} This work was supported by the Polish National Science Centre research project DEC-2011/02/A/HS1/00395.

\bibliographystyle{plain}

\begin{thebibliography}{16}
%
%
%
\bibitem{CanGolNic14}
D. Cantone, J. Goli\'{n}ska-Pilarek, M. Nicolosi-Asmundo.
\newblock A Relational Dual Tableau Decision Procedure for Multimodal and Description Logics.
\newblock To appear in: {\em Proceedings of the 9th International Conference on Hybrid Artificial Intelligence Systems}, Salamanca, Spain, 11th - 13th June 2014.
%
\bibitem{CanNicOrl10}
D. Cantone, M. Nicolosi Asmundo, E. Or{\l}owska.
\newblock Dual tableau-based decision procedures for some
relational logics.
\newblock In: {\em Proceedings of the 25th Italian Conference on Computational Logic}, Rende, Italy, July 7-9, 2010, pp. 1--16.
CEUR Workshop Proceedings vol. 598.
%
\bibitem{CanNicOrl11}
D. Cantone, M. Nicolosi Asmundo, E. Or{\l}owska.
\newblock Dual tableau-based decision procedures for relational logics with restricted composition operator.
\newblock {\em Journal of Applied Non-classical Logics} 21, No 2, 2011, 177-200.


\bibitem{DeOrVa99}
S. Demri, E. Orlowska, D. Vakarelov.
\newblock Indiscernibility and complementarity relations in information systems.
\newblock In: {\em J. Gerbrandy, M. Marx, M. de Rijke and Y. Venema (eds) JFAK. Essays Dedicated to Johan van Benthem on the Occasion of his 50th Birthday}, Amsterdam University Press, 1999.

\bibitem{FoNi06} A. Formisano and M. Nicolosi Asmundo.
\newblock An efficient relational deductive system for propositional non-classical logics.
\newblock \emph{Journal of Applied Non-Classical Logics}, vol. 16(3-4), pp. 367-408 (2006).

%
%

\bibitem{GHM13} J. Goli\'{n}ska-Pilarek, T. Huuskonen, E.
Munoz-Velasco, \newblock Relational dual tableau decision procedures
and their applications to modal and intuitionistic logics.  \newblock
{\em Annals of Pure and Applied Logics} vol. 165 (2), pp. 409-–427 (2014).

\bibitem{GoMu11a} J. Goli\'{n}ska-Pilarek, E. Munoz-Velasco, and A.
Mora.  \newblock Implementing a relational theorem prover for modal
logic K. \newblock {\em International Journal of Computer
Mathematics}, 88(9):1869--1884, 2011.

\bibitem{GoMu11b} J. Goli\'{n}ska-Pilarek, E. Munoz-Velasco, and A. Mora.
\newblock A new deduction system for deciding validity in modal logic K.
\newblock {\em Logic Journal of IGPL} 19(2):425--434, 2011.
%
\bibitem{GolOrl07} J. Goli\'{n}ska-Pilarek, E. Or{\l}owska.
\newblock Tableaux and dual tableaux:
Transformation of proofs.
\newblock {\em Studia Logica}, 85(3):283�-302, 2007.
%
%
%
\bibitem{Orl88} E. Or\l owska.
\newblock Relational interpretation of modal logics.
\newblock In: {\em H. Andreka, D. Monk, and I. Nemeti eds., Algebraic Logic. Colloquia
Mathematica Societatis Janos Bolyai}, vol. 54, pp. 443--471, North
Holland, 1988.
%
\bibitem{Orl} E. Or{\l}owska, J. Goli\'{n}ska-Pilarek.
\newblock Dual Tableaux: Foundations,
Methodology, Case Studies.
\newblock {\em Trends in Logic} vol. 36, Springer, 2011.
%
%
\bibitem{TarskiGivant87} A. Tarski, S. Givant.
 \newblock A Formalization of Set Theory without Variables.
 \newblock {\em American Mathematical Society Colloquium Publications}, Providence, Rhode Island, 1987.
\end{thebibliography}

\end{document}